\documentclass[12pt,3p,preprint,sort&compress]{elsarticle}
\usepackage[colorlinks=true]{hyperref}

\usepackage{amsmath}
\usepackage{amssymb}
\usepackage{mathtools}
\usepackage[utf8]{inputenc}

\newcommand{\e}{{\mathrm{e}}}

\renewcommand{\d}{\partial}
\renewcommand{\l}{\left(}
\renewcommand{\r}{\right)}
\newcommand{\la}{\langle }
\newcommand{\ra}{\rangle }
\newcommand{\be}{\begin{equation}}
\newcommand{\ee}{\end{equation}}
\newcommand{\ba}{\begin{align}}
\newcommand{\ea}{\end{align}}
\newcommand{\bg}{\begin{gather}}
\newcommand{\eg}{\end{gather}}
\newcommand{\bseq}{\begin{subequations}}
\newcommand{\eseq}{\end{subequations}}

\newcommand{\half}{\frac{1}{2}}

\begin{document}
\begin{flushright}
	INR-TH-2025-001
\end{flushright}

\title{Playing with lepton asymmetry at the resonant production of sterile neutrino dark matter} 
\author[inr,mpti]{Dmitry Gorbunov}
\ead{gorby@ms2.inr.ac.ru}
\author[inr,mpti]{Dmitry Kalashnikov}
\ead{kalashnikov.d@phystech.edu}
\author[inr,mpti]{George Krugan}
\ead{gnu271428@gmail.com}
\address[inr]{Institute for Nuclear Research of Russian Academy of Sciences, 117312 Moscow, Russia}
\address[mpti]{Moscow Institute of Physics and Technology, 141700 Dolgoprudny, Russia} 
\begin{abstract}
    We examine the sterile neutrino dark matter production in the primordial plasma with lepton asymmetry unequally distributed over different neutrino flavors. 
    We argue that with the specific flavor fractions, one can mitigate limits from the Big Bang Nucleosynthesis on the sterile-active neutrino mixing angle and sterile neutrino mass. It happens due to cancellation of the neutrino flavor asymmetries in active neutrino oscillations, which is more efficient in the case of inverse hierarchy of active neutrino masses and does not depend on the value of CP-phase. 
    This finding opens a window of lower sterile-active mixing angles. Likewise, we show that, with lepton asymmetry disappearing from the plasma at certain intermediate stages of the sterile neutrino production, the spectrum of produced neutrinos becomes much colder, which weakens the limits on the model parameter space from observations of cosmic small-scale structures (Ly-$\alpha$ forest, galaxy counts, etc.). This finding reopens the region of lighter sterile neutrinos. The new region may be explored with the next generation of X-ray telescopes searching for the inherent peak signature provided by the dark matter sterile neutrino radiative decays in the Galaxy.  
\end{abstract}
\date{}

\maketitle


{\bf 1.} Sterile neutrino, massive fermion, singlet under the Standard Model gauge group, is one of the physically well-motivated candidates to serve as dark matter. Sterile neutrinos are naturally introduced as an essential part of the see-saw type I mechanism possibly responsible for active neutrino masses \cite{Schechter:1980gr}, which are required to explain neutrino oscillations. The mechanism yields mixing between active and sterile neutrinos making the latter unstable. However, with sufficiently small mixing, sufficiently light sterile neutrinos (typically in keV mass region) are long-lived enough to entirely populate the present dark matter component \cite{Dodelson:1993je,Boyarsky:2009ix,Abazajian:2017tcc}. 

The active-sterile neutrino mixing plays a two-fold role in the sterile neutrino cosmology. First, it ensures sterile neutrino production via  active-sterile neutrino oscillations in the primordial plasma \cite{Dodelson:1993je,Shi:1998km}. Second, the mixing induces sterile neutrino radiative decay \cite{Pal:1981rm}, decay into active neutrino and photon, which provides a peak signature to be searched for in the Galactic X-ray spectrum. Naturally, the absence of the peak signature places an upper limit on the sterile-active neutrino mixing angle $\theta$, the strongest one comes from observations of XMM \cite{Foster:2021ngm} and NuSTAR\,\cite{Krivonos:2020qvl,Roach:2022lgo,Krivonos:2024yvm} telescopes. 

However, $\theta$ may also be limited from below if the only source of sterile neutrino DM production is the sterile-active mixing \footnote{There are many models where other interactions are producing DM sterile neutrinos \cite{Shaposhnikov:2006xi,Kusenko:2006rh,Bezrukov:2011sz,Shaposhnikov:2020aen,Dror:2020jzy}.}. This is the case we study below. 

It is well known \cite{Shi:1998km} that the efficiency of sterile neutrino production grows with the amount of lepton asymmetry in the plasma. The change of neutrino dispersion relation in the asymmetric plasma leads to the resonant amplification of the active neutrino transitions to the sterile component.  

The lepton asymmetry may influence the later cosmological stages, in particular,  primordial nucleosynthesis and recombination. Observations of primordial helium abundance favor standard cosmology 
and severely constrain \cite{ParticleDataGroup:2024cfk,Escudero:2022okz} the asymmetry in electron neutrinos at the moment of neutron freeze-out, which happens at the plasma temperature of about $T_n\approx 0.7$\,MeV, some  time after neutrino decoupling. As we show below, the lower bounds on mixing $\theta$,  inferred from this constraint, can be significantly reduced in the case of a specific initial lepton asymmetry pattern, with large individual flavor asymmetries but a total lepton asymmetry close to zero. Then active neutrino oscillations, which start at temperature $T\lesssim 20$\,MeV, very effectively cancel all flavor asymmetries in accordance with calculations of Refs.\,\cite{Dolgov:2002ab,Barenboim:2016shh}. We find that the cancellation is more efficient in the case of inverse ordering of neutrino masses, and CP-phase does not play a noticeable role. This allows the larger flavor lepton asymmetries at sterile neutrino production, hence lower mixing angles, consistent with observed helium abundance and anisotropy of the cosmic microwave background.  

The sterile neutrino mass $m_{\nu_s}$ is also bounded from below by observations of the cosmic structure formation. Being produced in oscillations of the active neutrino, the sterile neutrino inherits its momentum, which is generically of order of the plasma temperature. Hence, due to redshifting, a keV-scale sterile neutrinos exhibit velocities of order $10^{-3}$ at the onset of cosmic structure formation, at the epoch of equality between radiation and matter, when the plasma temperature is $T\approx 0.7$\,eV. The dark matter candidates with this feature are called warm dark matter. The matter inhomogeneities at short distances are smoothed off by the free streaming of such dark matter particles. Observations of dwarf galaxies constrain the warm dark matter models with too high velocities, in particular, exclude too light sterile neutrinos, for recent studies see e.g. \cite{Zelko:2022tgf,Newton:2024jsy,Bezrukov:2024qwr}. We show below, that this constraint can be weakened if the lepton asymmetry, invoked in the sterile neutrino production, disappears  in the early Universe thus terminating the process. The reason is that the resonant production starts with particles of lower, than average, momenta and later proceeds to particles of higher momenta. Hence, if the lepton asymmetry disappears in the middle of the process, the final spectrum of the produced neutrinos turns colder than normally expected. Consequently, the average velocity of sterile neutrinos at the equality becomes lower, and then smaller sterile neutrino masses become consistent with the cosmic structure formation.       

We explain both options with an analysis of a simple Boltzmann equation describing the neutrino oscillations in plasma. We illustrate the expected results with the help of numerical codes \texttt{dmpheno} \cite{Ghiglieri:2015jua} and \texttt{FortEPiaNO} \cite{Gariazzo:2019gyi} ( which we slightly modified to suit our purposes). The first code evaluates the sterile neutrino production in a model, where all three neutrino asymmetries are taken into account. The second code traces the evolution of active neutrino asymmetries in the early Universe from the onset of active neutrino oscillations in the primordial plasma till the active neutrino decoupling.

{\bf 2.} 
The out-of-equilibrium production of sterile neutrinos $\nu_s$ in oscillations of active neutrinos $\nu_\alpha$, $\alpha=e,\,\mu,\,\tau$, can be described by the Boltzmann equation 
\begin{equation}
    \label{Boltz}
\frac{\d f_{\nu_s}}{\d t} - H p \frac{\d f_{\nu_s}}{\d p} = 
\frac{\Gamma_{\nu_\alpha}(p)}{2} \la P(\nu_\alpha\to\nu_s)\ra f_{\nu_\alpha} 
\end{equation}
where $f_\nu(p,t)$ is neutrino $\nu$ distribution over 3-momentum $p$, $H(t)$ is the Hubble parameter, the expansion rate of the Universe, $\Gamma_{\nu_\alpha}$ is the neutrino $\nu_\alpha$ weak interaction rate in the plasma, which is of order $G_F^2T^5$ with Fermi constant $G_F$ and plasma temperature $T$. It is also assumed that the weak interaction rate is much lower than the rate of oscillations between the active and sterile neutrinos, i.e. 
\[
G_F^2T^5\sim \Gamma_{\nu_\alpha}(p) \ll \frac{m_{\nu_s}^2}{2p}\sim \frac{m_{\nu_s}^2}{2T}\,, 
\]
so the oscillation probability is obtained upon averaging over several oscillation periods \cite{Venumadhav:2015pla}
\begin{equation}
    \label{probability}
 \la P(\nu_\alpha\to\nu_s)\ra  =  \half \,\frac{\frac{m_{\nu_s}^4}{4p^2} \sin^22\theta_\alpha}{ \frac{m_{\nu_s}^4}{4p^2} \sin^22\theta_\alpha + \frac{\Gamma_{\nu_\alpha}^2}{4}+ \l \frac{m_{\nu_s}^2}{2p} \cos2\theta_\alpha - V^L_\alpha-V^T_\alpha\r^2}\,. 
\end{equation}
Here we introduced the neutrino thermal potential in the plasma, which is negative \cite{Venumadhav:2015pla}, 
\[
V^T_\alpha(p)=-\frac{8\sqrt{2}}{3}\,G_F\,\l \frac{\rho_{\nu_\alpha}}{M_Z^2} +\frac{\rho_\alpha}{M_W^2}\r p\,,
\]
with $Z$-boson mass $M_Z$, $W$-boson mass $M_W$, energy density of neutral $\rho_{\nu_\alpha}$ and charged $\rho_\alpha$ leptons in the plasma, 
and lepton asymmetry potential \cite{Venumadhav:2015pla}
\[
V^L_\alpha\!=\! \sqrt{2} G_F\! \left[\! \l 1\!-\!2 \sin^2\theta_W\r\!\Delta n_Q -\half \Delta n_B\!+\!\sum_\beta\!\l \l 1\!+\!\delta_{\alpha\beta}\r \Delta n_{\nu_\beta}\r +\l \!\delta_{\alpha\beta}\!-\!\half\!+\!2\sin^2\theta_W\!\r\!\Delta n_\beta\right]
\]
with number densities of baryon charge $\Delta n_B$, electric 
charge $\Delta n_Q$, neutral lepton asymmetries 
$\Delta n_{\nu_\beta}$ and charged lepton asymmetries $\Delta n_\beta$, see e.g. \cite{Venumadhav:2015pla} for details. 
Hereafter we loosely use the same notation $p$ for neutrino momentum and neutrino energy, since all neutrinos are ultra-relativistic during the active period of sterile neutrino production, starting from temperature of several GeV to dozens of MeV. 

Note that the production rate of sterile neutrinos grows proportionally to the weak interaction rate, see r.h.s. of eq.\,\eqref{Boltz}, 
since interaction in plasma collapses the neutrino wave function, resulting in the appearance of a sterile neutrino with probability \eqref{probability}. The weak rate is higher at higher temperature, but the transition probability \eqref{probability} is highly suppressed both by damping (the second term in the denominator) and by thermal potential. Hence, generically the oscillation probability in plasma is lower than in vacuum, but for the lepton asymmetry potential which may cancel the mass term in the brackets and thus amplifies the transition similarly to the Mikheev--Smirnov--Wolfenstein effect in the Sun. The lepton asymmetry potential also decreases with temperature, $V^L_\alpha\propto T^3$,  but slower than the thermal potential, $V^T_\alpha\propto T^5$. Therefore, the oscillation probability is maximized when the quantity in brackets is close to zero, 
\begin{equation}
\label{resonance}
\frac{m_{\nu_s}^2}{2p} \cos2\theta_\alpha - V^L_\alpha-V^T_\alpha\approx 0\,,
\end{equation}
that is possible to achieve with sufficiently large (as compared to the baryon asymmetry) lepton asymmetry. Actually, the resonance condition \eqref{resonance} is a quadratic equation with respect to momenta (since $V^T_\alpha\propto p$), and potential $V^L(T)$ allows for possible physical solutions of positive $p$ (recall that $V^T_\alpha$ is negative). 

To analyze the expected spectrum  analytically, it is useful to cast the Boltzmann equation in terms of variables $T$ and $y\equiv p/T$  and further neglect the decoupling of relativistic degrees of freedom from the plasma in the interesting epoch. Then l.h.s. of the Boltzmann equation \eqref{Boltz} transforms into $HT\d f/\d T$, while the resonance condition \eqref{resonance} may be written as follows 
\begin{equation}
    \label{quadratic}
    \frac{A}{y}\frac{m^2_{\nu_s}}{M_W^2} - B  \frac{T^4}{M_W^4}\Delta_{\nu_\alpha} +C \frac{T^6}{M_W^6} y =0\,,
\end{equation}
where the positive dimensionless parameters $A$, $B$, $C$ (not necessary of order one) do not depend on the variables $y$, $T$ and can be directly extracted from the above formulas;  dimensionless parameter $\Delta_{\nu_\alpha}\equiv \Delta n_{\nu_\alpha}/s$,  with entropy density $s=2\pi^2 h_{eff} T^3/45$\,\cite{Rubakov:2017xzr} of plasma with $h_{eff}$ degrees of freedom, characterizes the neutrino flavor asymmetry (the lepton flavor asymmetry is roughly proportional to it with the factor of about 3). Naturally, there are two solutions of the quadratic equation \eqref{quadratic}. Since the second and third terms exhibit strong dependence on temperature, one can consider the regimes of (relatively) large and small temperatures. In the first regime the solution is mostly saturated by the second and third  terms in \eqref{quadratic}, so the resonance happens for the particles of momentum $p=yT$  with 
\begin{equation}
\label{first}
    y\sim y_{r1}\equiv \Delta_{\nu_\alpha}\frac{B}{C}\,\frac{M_W^2}{T^2}\,.
\end{equation}
In the second regime the first and second terms are important, and the resonance develops for particles of momentum $p=yT$ with 
\begin{equation}
\label{second}
    y\sim y_{r2}\equiv\frac{1}{\Delta_{\nu_\alpha}}\frac{A}{B}\,\frac{m^2_{\nu_s}}{M_W^2}\frac{M_W^4}{T^4}\,.
\end{equation}

\begin{figure}[!htb]
    \centerline{
    \includegraphics[width=0.33\textwidth]{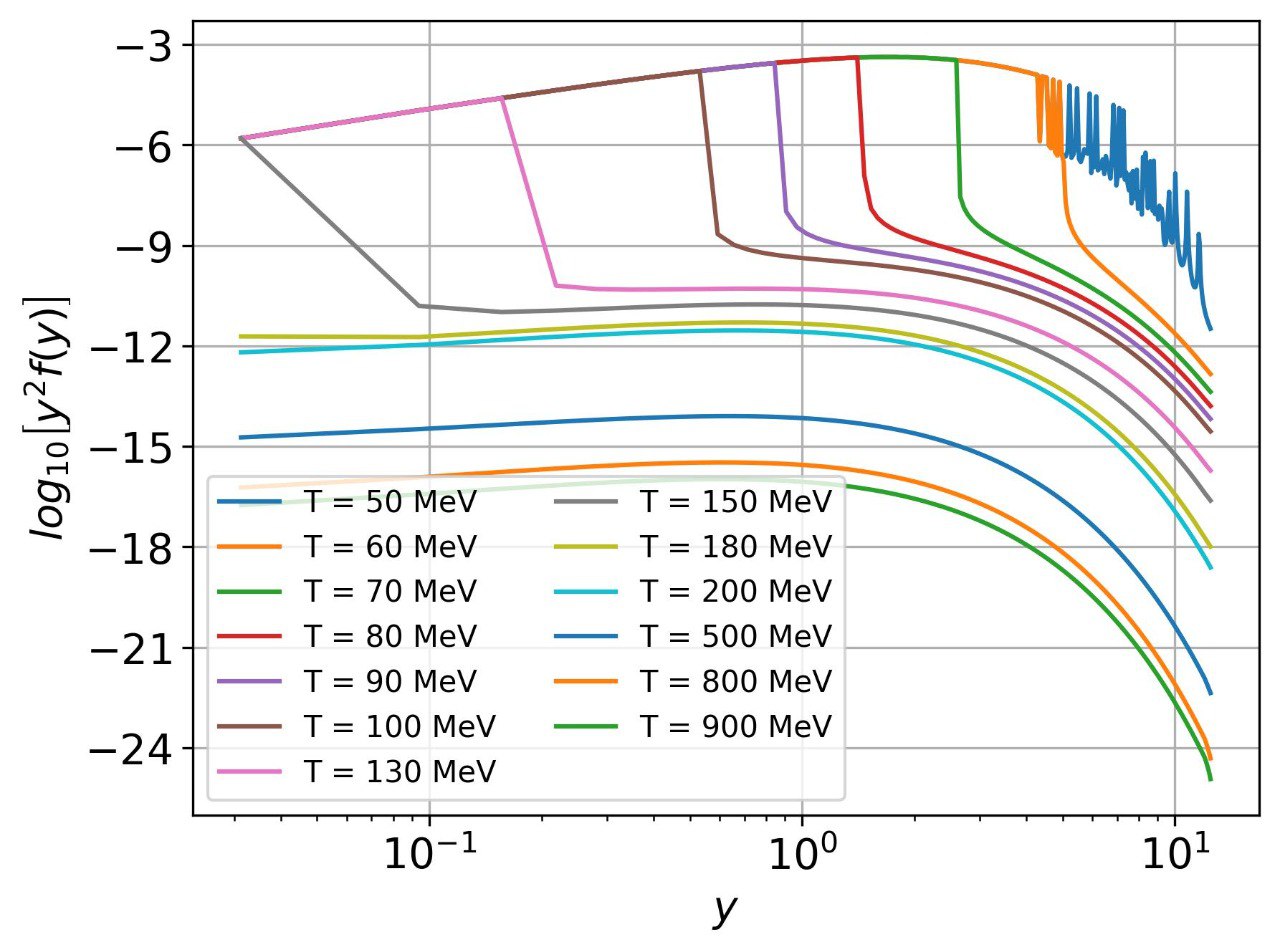}
    \includegraphics[width=0.33\textwidth]{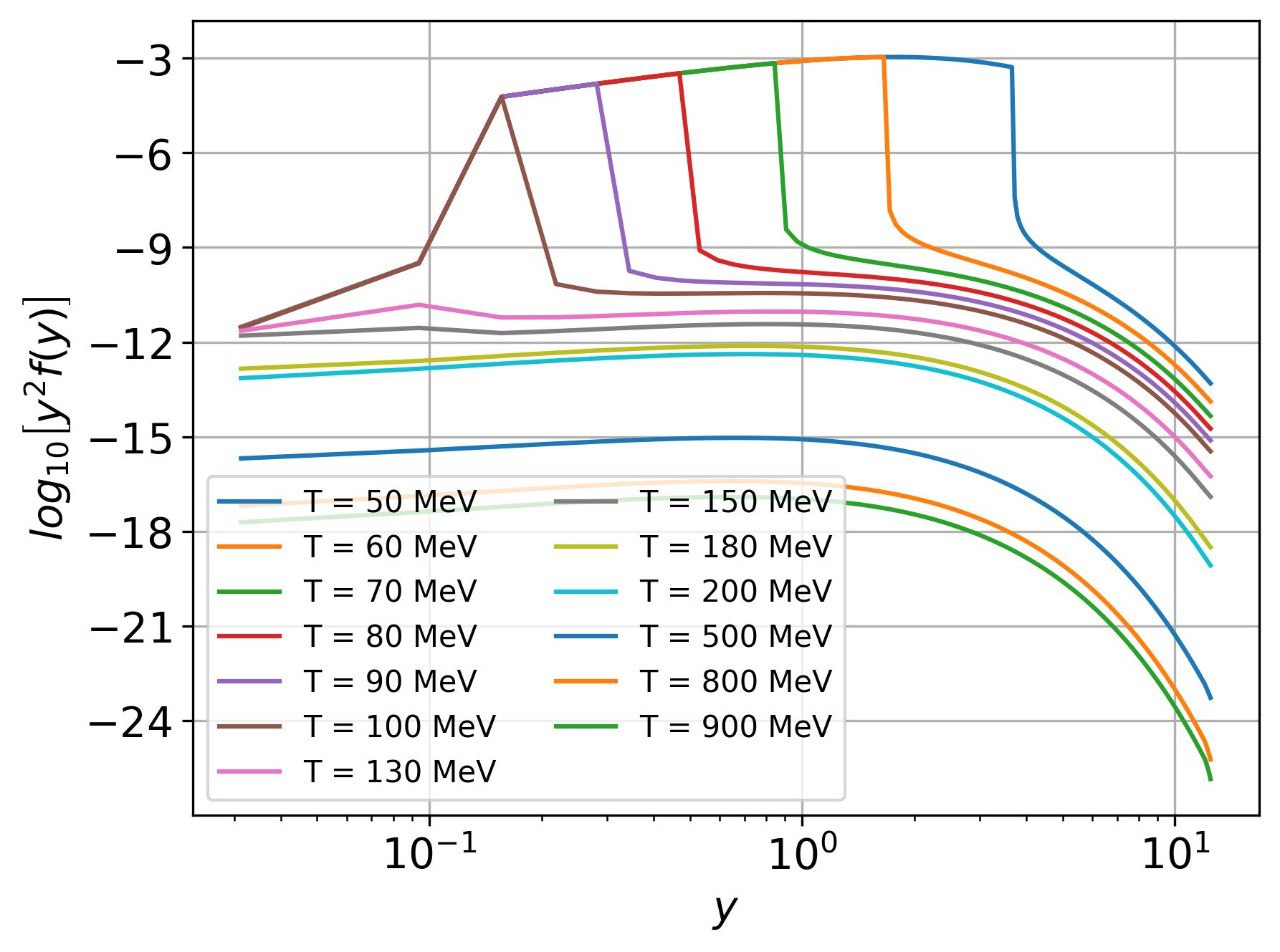}
    \includegraphics[width=0.33\textwidth]{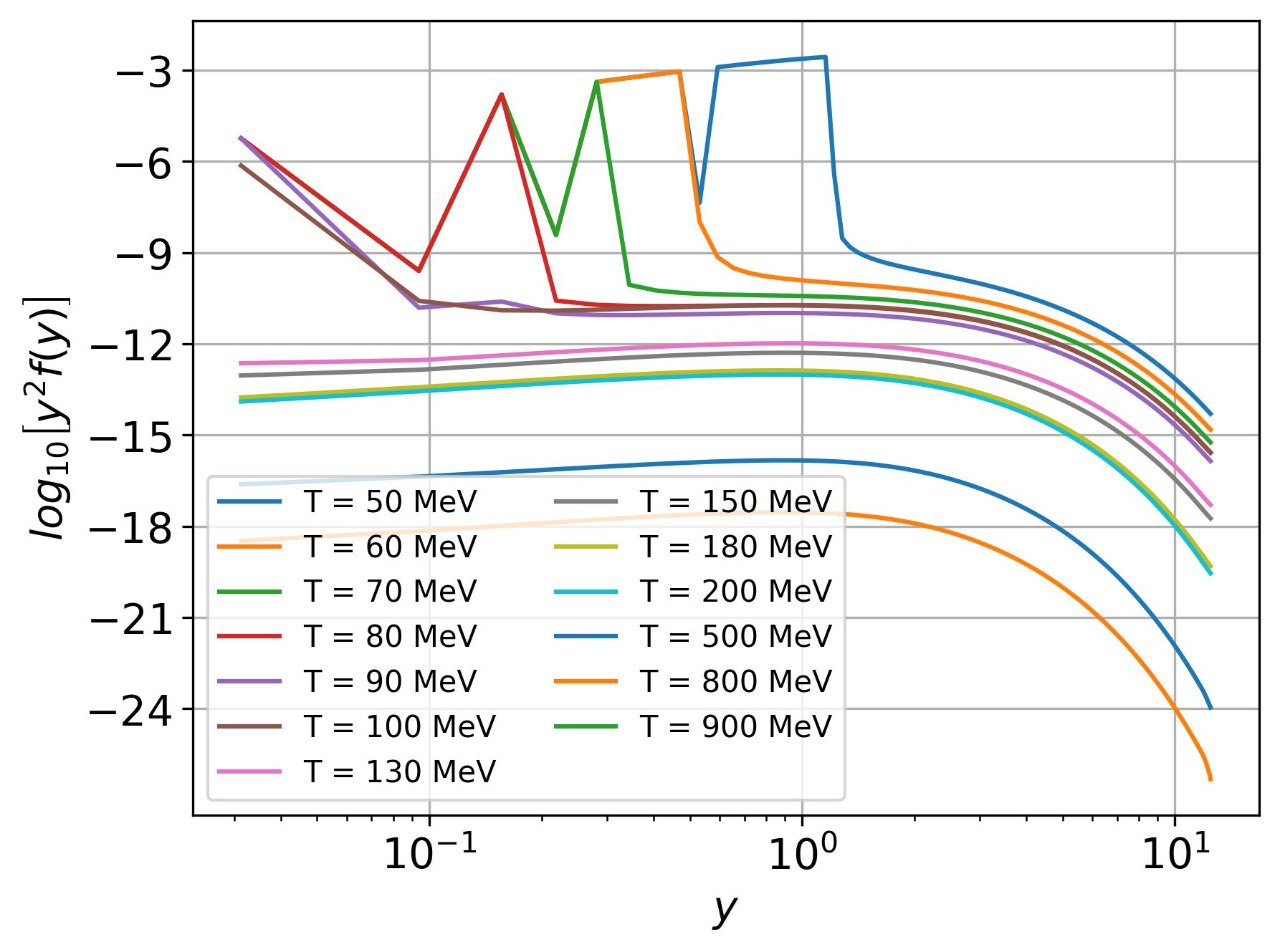}
    }
    \caption{Spectra of sterile neutrinos produced via mixing with electron neutrinos, $y=p/T$, at various plasma temperatures $T$ and various initial lepton asymmetries. The initial flavor neutrino asymmetries are equal to $(1,\,-0.5\,-0.5)\cdot N\cdot 10^{-3}$ for electron, muon and tau. The initial values of the scale factor $N$ used in calculations for the three plots are $1,\,3\,,9$ from left to right. The dips in the spectrum are caused by the limited precision of the numerical calculations, which could be improved but at the cost of significantly increased computational time. }
    \label{fig:spectra}
\end{figure}

First, one observes that in both cases \eqref{first}, \eqref{second} the resonance production starts with the coldest particles,  because the resonance parameters $y_{r1}$, $y_{r2}$ grow with decreasing temperature. The ordering of the solutions \eqref{first}, \eqref{second} depends on the dimensionless factor $\Delta_{\nu_\alpha} M_W/m_{\nu_s}$. Since in \eqref{probability} the damping term $\Gamma_{\nu_\alpha}^2\propto G_F^{4}T^{10}$  suppresses transitions at high temperatures, it is the second solution \eqref{second} that matters in the production of sterile neutrinos. It begins to produce relatively cold sterile neutrinos and later proceeds to produce the warmer ones. This behavior is illustrated in Fig.\,\ref{fig:spectra} 
by the spectrum calculation performed with the help of routine \texttt{dmpheno} \cite{Ghiglieri:2015jua}, the evolution starts from $T=4$\,GeV and we consider the model with sterile neutrino mass  $m_{\nu_s}=7.11$\,keV and mixing with only electron neutrino described by $\sin^22\theta=2.8\cdot 10^{-13}$. If the lepton asymmetry disappears, the second term in \eqref{quadratic} vanishes, that ceases the sterile neutrino production. The final spectrum of the produced sterile neutrino is colder in this case, and hence lighter sterile neutrinos become consistent with the cosmic structure formation. 
 Let us clarify that during the resonant oscillations some of the lepton asymmetry gets transferred to the sterile neutrinos. In the interesting region of tiny sterile-to-active mixing and relatively large initial lepton asymmetry, this transfer solely determines the final sterile neutrino abundance. However, numerically, the transfer is negligibly small, less than $10^{-4}$ of the initial lepton asymmetry. Moreover, we checked that we get the same numerical results while keeping the asymmetry constant at the evaluation. So, what we need above to cease the sterile neutrino production is  a disappearance of the major part of the asymmetry.

Second, the resonance momentum becomes colder with higher asymmetry $\Delta_{\nu_\alpha}$. It is confirmed in Fig.\,\ref{fig:spectra} 
with numerical calculations for three different initial asymmetries. At the same temperature, the spectra are colder for higher asymmetries (from left to right on the plot) in full accordance with \eqref{second}. Hence, exploiting higher asymmetry also helps to avoid the problems with cosmic structure formation and {\it reopens a window of light sterile neutrinos.} Moreover, with higher asymmetry the production starts later, reaching the epoch of active neutrino oscillations, $T\gtrsim 20$\,MeV, which very efficiently cancel asymmetry between the neutrino flavors (see below) hence finishing the sterile neutrino production and consequently {\it yielding generically colder spectrum.}

Alternatively, one may reduce the asymmetry and stop the production by involving some new ingredients. There is an issue here. 
Though, the presented in Fig.\,\ref{fig:spectra} spectra clearly show what can be achieved with instant disappearance of the lepton asymmetry, in realistic mechanisms the operation takes some time, and so the spectra are smoother than one could envisage from the plots. 
Actually, by reducing the asymmetry one shifts the most effective production to earlier stages but makes the production longer, and so the average momentum remains almost intact. We observed this behavior in a realistic situation with one more, but much heavier sterile neutrino component in the plasma. We assume, that these neutrinos are non-relativistic and their decay rate is about the Hubble parameter in the interesting epoch, $\Gamma_d=H(T_d)$. The decay products are asymmetric in neutrino flavors and tend to reduce the neutrino asymmetries. This reduction changes the final spectra of the sterile neutrino very little, while we vary the temperature in the interval  $80<T_d<260$\,MeV. 

\begin{figure}[!htb]
    \centerline{
    \includegraphics[width=0.33\textwidth]{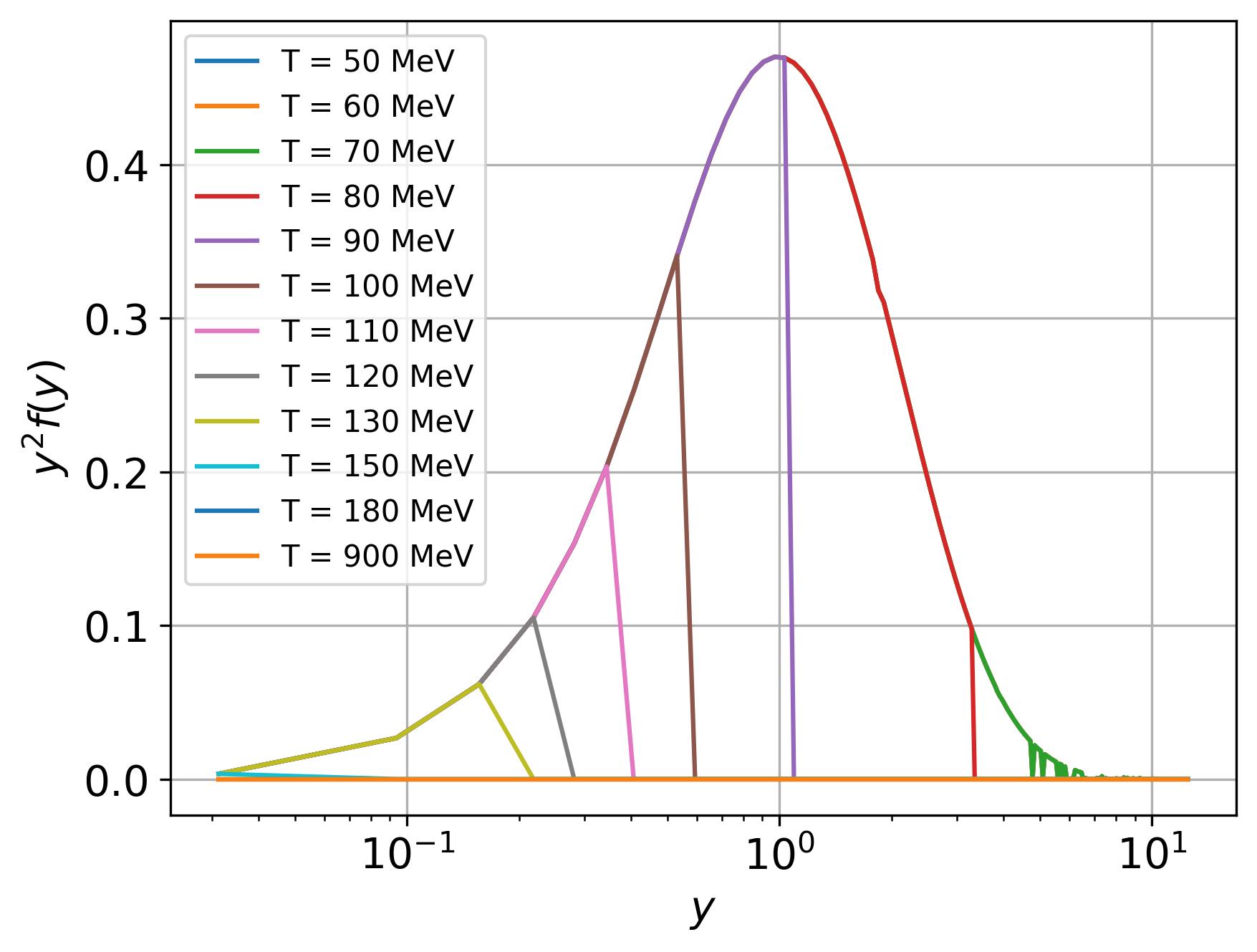}
    \includegraphics[width=0.33\textwidth]{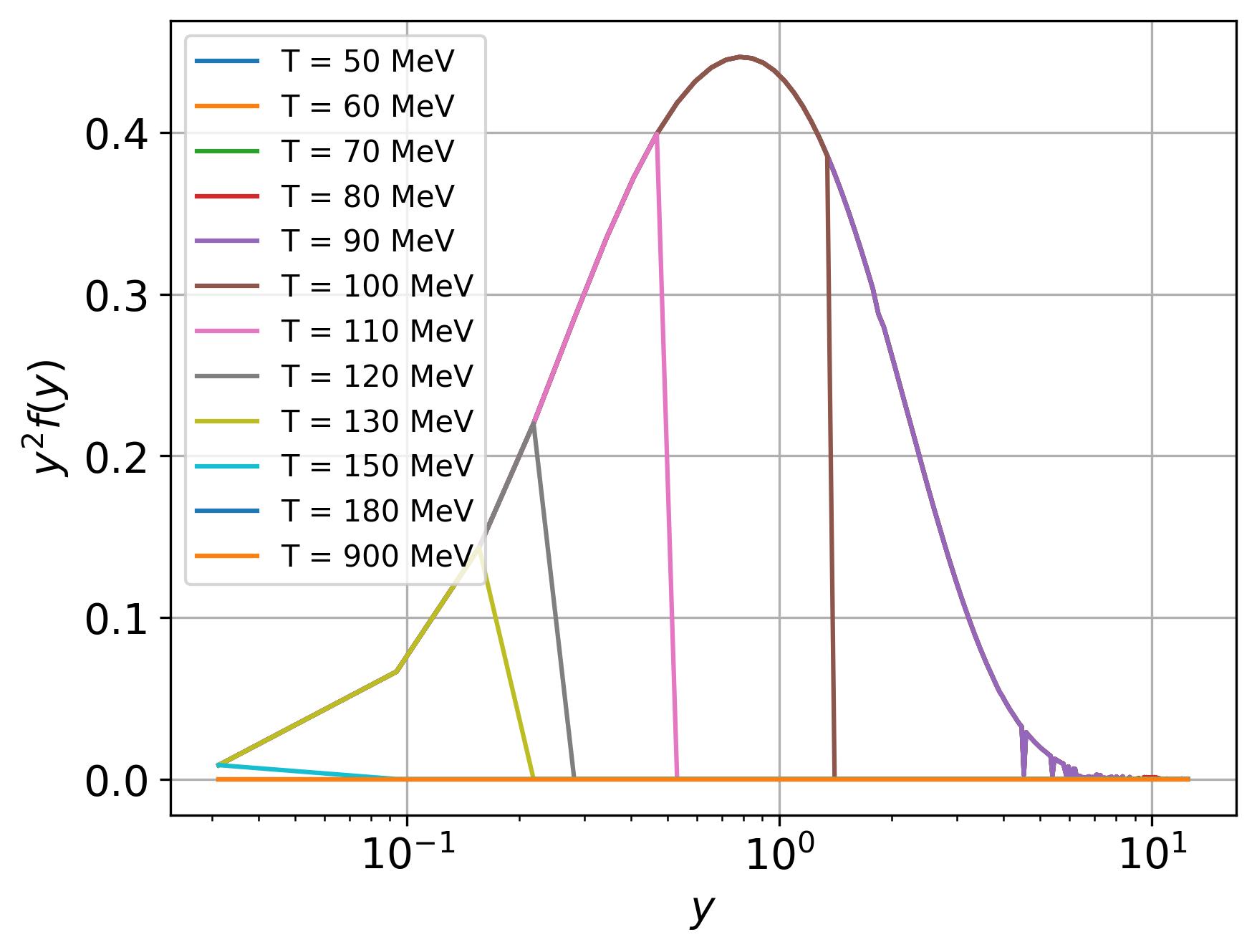}
    \includegraphics[width=0.33\textwidth]{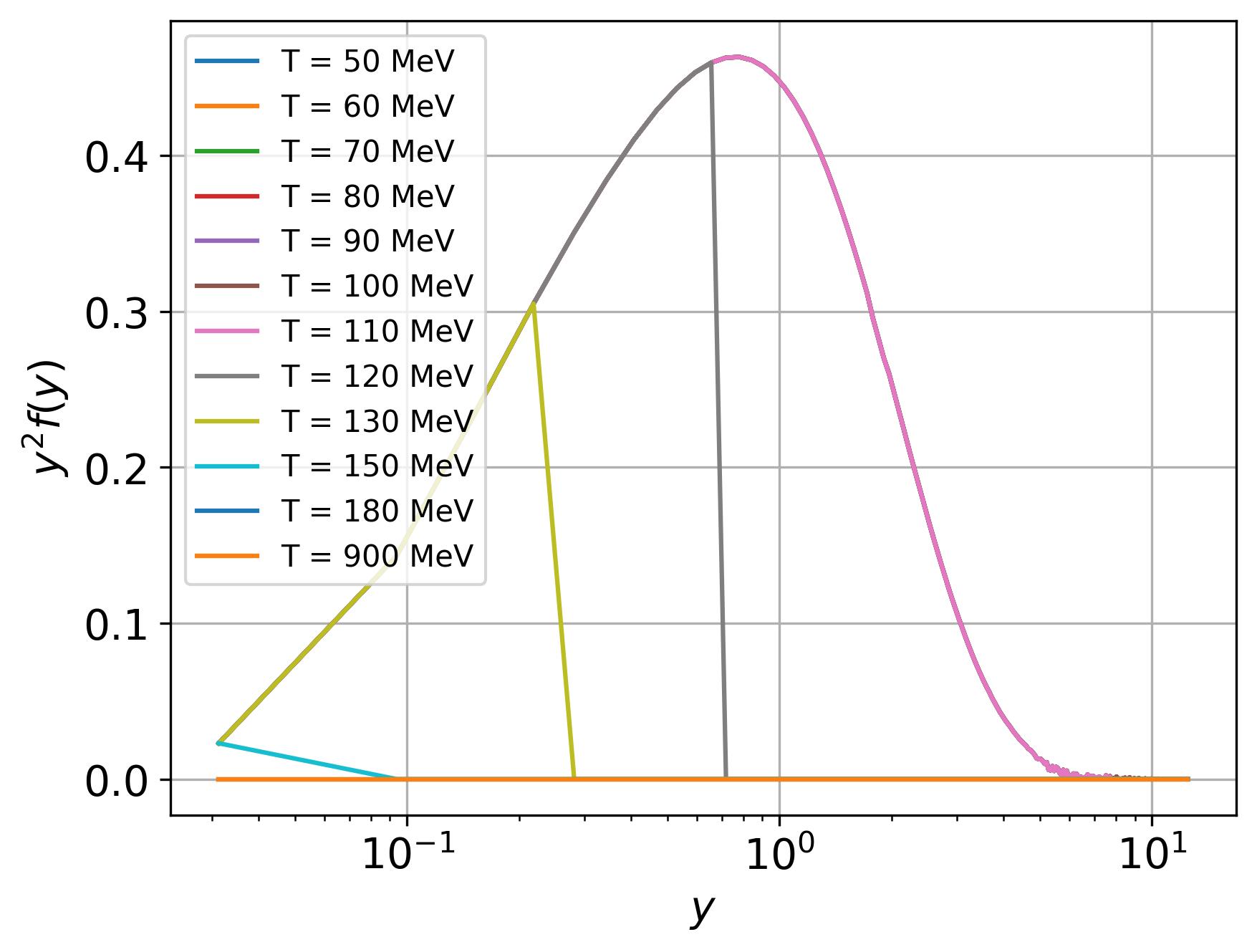}
    }
    \caption{Normalized to unity spectra of sterile neutrinos produced via mixing with electron neutrinos, $y=p/T$, at various plasma temperatures $T$ and initial lepton asymmetries $(1,\,-0.5\,-0.5)\cdot 10^{-3}$ for electron, muon and tau. 
    The asymmetries are reduced because of flavor violating  heavy neutral fermions decays of heavy neutral fermions in plasma with decay rates $\Gamma_d=H(T_d)\times (T_d/T)^{10}$ and reference temperatures $T_d$ (from left to right): 80\,MeV, 100\,MeV, 120\,MeV. }
    \label{fig:reduction}
\end{figure}

However, the spectrum can be deformed with close to instant change of the asymmetry, if the change rate much exceeds the Universe expansion rate. Such a situation is natural in a specific but well-motivated models, e.g. the QCD axion mass grows fast and becomes much higher than the Hubble rate during the QCD transition. 
As an example, one may suggest that the fermion mass depends on some cosmic scalar field, which evolves linearly with cosmic time. The fermion decay rate into active neutrinos, induced by mixing and weak interaction, then strongly depends on the cosmic time $\Gamma_d\propto t^5$. To mimic this situation, we added to the equations for the asymmetries in the code \texttt{dmpheno} \cite{Ghiglieri:2015jua} the new terms which induce the asymmetries at the rate $\Gamma=H(T_d)\times (T/T_d)^{10}$. We calculated the sterile neutrino production for three values $T_d=80,\,100,\,120$\,MeV and the same initial neutrino asymmetries as we used for the left plot in Fig.\,\ref{fig:spectra}. We present the obtained spectra in Fig.\,\ref{fig:reduction}.  
One can see that the maxima of the final spectra shift towards the cold direction with earlier destruction of the lepton asymmetry (from possible decays of heavy fermions). There is no need to make the asymmetry equal to zero, just to reduce it significantly, e.g. for the right plot in Fig.\,\ref{fig:reduction} the final neutrino asymmetries are $(-1.93,\,-0.04,\,3.75)\cdot 10^{-4}$. Reducing the asymmetry in this manner shifts the resonant production toward higher values of y, which results in a warmer spectrum. This effect competes with the cooling effect of the decays, which consequently limits the reduction in the mean momentum $p = yT$).


{\bf 3.} A larger lepton asymmetry leads to stronger amplification of the resonance production. One can check that at large asymmetry the amount of produced particles scales linearly with asymmetry. Thus, the larger the asymmetry, the lower the mixing angle $\theta$ is needed to provide the same amount of dark matter. However, when oscillations into sterile neutrino terminate, the net lepton asymmetry remains in plasma. In the literature, typically, it is supposed that the asymmetry resides mostly in one neutrino flavor. Then, the active neutrino oscillations in plasma, which begin at temperature of about 20\,MeV, redistribute it over all three species. Then the electron neutrino asymmetry approaches one third of the initial asymmetry and must obey the upper limit inferred from observations of the helium abundance\,\cite{ParticleDataGroup:2024cfk,Escudero:2022okz}. It implies a lower limit on the mixing angle $\theta$ consistent with the hypothesis that sterile neutrinos, generated in this way, fully explain the dark matter component. 

This limit may be significantly alleviated if the {\it initial lepton asymmetry}, used in the dark matter generation, is distributed between all three flavors in such a way that each flavor asymmetry is larger in value, but the total lepton asymmetry is close to zero. In such a situation, the active neutrino oscillations tend to reduce each of the flavor asymmetry, including the electron neutrino asymmetry, to zero, see Refs.\,\cite{Dolgov:2002ab,Barenboim:2016shh}. 

Potentially, this observation allows one to use any large asymmetry at the beginning of sterile neutrino production as far as the sum of final lepton flavor asymmetries is close to zero. In practice, the lepton asymmetry redistribution is quite limited in time. It starts with the onset of oscillations at a temperature of about 20\,MeV and stops with neutrino decoupling from the primordial plasma, when its temperature drops to about 3\,MeV. 

The redistribution of the asymmetry between the flavors at the fast oscillation stage is very natural, because the flavor changing processes are in equilibrium, while the total lepton number is conserved. It can be illustrated by considering oscillations between two species, say $\nu_\tau$ and $\nu_\mu$, where the redistribution process minimizes the difference $L_{\tau\mu}\equiv \Delta n_{\nu_\tau}-\Delta n_{\nu_\mu}$. The evolution of neutrino distributions over momenta $f_{\nu_\alpha}$ can be described with the simplified Boltzmann equations, e.g. 
\begin{equation}
\label{B-as}
\frac{\d f_{\nu_\tau}}{\d t} - H p \frac{\d f_{\nu_\tau}}{\d p} = 
\frac{\Gamma_w(p)}{2} \la P(\nu_\mu\to\nu_\tau)\ra \l  f_{\nu_\mu}-f_{\nu_\tau}\r 
\end{equation}
and similar for antineutrinos. Here only neutral currents mediate the weak interactions, and so the weak rate $\Gamma_w$ is the same for all equations. As in eq.\,\eqref{Boltz} we assume that the oscillations are faster than the weak interactions, but also account for the inverse oscillation processes. We neglect all the processes not participating in the asymmetry redistribution, e.g. weak scatterings, which keep the distribution functions of the thermal form. 
The interesting asymmetry difference is determined by 
\[
L_{\tau\mu}\propto \int d^3p \l f_{\nu_\tau}-f_{\bar\nu_\tau} -f_{\nu_\mu}+f_{\bar\nu_\mu}\r\,.
\]
Then forming a relevant combination by summing up the Boltzmann equations\,\eqref{B-as} and integrating over 3-momenta (one then replaces the momenta in $\Gamma_w$ by its thermal average) we obtain the equation for the asymmetry difference
\[
\frac{dL_{\tau\mu}}{dt}+3HL_{\tau\mu}=-D\cdot\Gamma_{w}(T)\, L_{\tau\mu}\,,
\]
where 
coefficient $D$ depends on the entries of neutrino mixing matrix. 

This is an approximate equation but it captures all the main ingredients of the asymmetry distribution process in general case. 
Omitting the indices, 
replacing cosmic time $t$ with temperature $T$ and substituting the Hubble parameter as a function of temperature and Planck mass $M_{Pl}$, we find for the asymmetry-to-entropy ratio, $\Delta\equiv L/s$, 
\begin{equation}
\label{washing}    
\Delta(T)=\Delta(T_i)\,\e^{F\cdot G_F^2 M_{Pl}\l T^3-T_i^3\r},
\end{equation}
where $F$ stands for some numerical factor. We observe that the exponential factor is determined by the temperature $T_i$ at the onset of oscillations, which occurs when the oscillation rate exceeds the weak interaction rate. More accurately, we are interested in the moment, when the 
vacuum term overcomes the matter term in the effective neutrino Hamiltonian. It implies, for average neutrino momenta, 
\[
\frac{\Delta m^2}{2 T}\gtrsim \varkappa G_F^2T^5\,,
\]
hence, 
\[
T_i\simeq \l \varkappa^{-1}G_F^{-2}\Delta m^2 \r^{1/6}\,,
\]
with flavor-dependent numerical factor $\varkappa$. 

The temperature at which the oscillations start depends on the difference $\Delta m^2$ of squared neutrino masses. Experimental investigations of neutrino oscillations reveal two squared mass differences \cite{ParticleDataGroup:2024cfk}, 
\[
\Delta m_{sol}^2\approx 7.5\cdot 10^{-5}\,\text{eV}^2\,,\;\;\;\;
\Delta m_{atm}^2\approx 2.5\cdot 10^{-3}\,\text{eV}^2\,.
\]
They are hierarchical, and so in the expanding Universe first start oscillations, which cancel neutrino asymmetries between tau and muon neutrino flavors. It occurs at a temperature of about 
\[
T_{i,\mu\tau}\approx 20\,\text{MeV}\,.
\]
At this and lower temperatures the electron neutrinos interact in the plasma somewhat stronger than the muon and tau neutrinos, due to the presence of electron and positron pairs in the plasma, which yields $\varkappa_e/\varkappa_{\mu,\tau}\approx 4$\,\cite{Dolgov:2002ab}.  
Oscillations, which engage the electron 
neutrino asymmetry in this process, start at a lower temperature of about 
\[
T_{i,e}\sim T_{i,\mu\tau} \times \l\frac{\varkappa_\tau\Delta m_{sol}^2}{\varkappa_e\Delta m_{atm}^2}\r^{1/6}\simeq 0.4\,T_{i,\mu\tau}\simeq 8\,\text{MeV}\,.
\]
The cubic dependence on temperature implies at least an order of magnitude  smaller numerical factor (as compared to the factor working for other two flavors) in the exponent \eqref{washing}. Therefore, at the decoupling of neutrino from plasma, at the temperature of about $T_d\simeq 3$\,MeV, the expected asymmetries would be 
\[
\Delta_{\mu\tau}(T_d)=\Delta_{\mu\tau}(T_{i,\mu\tau})\e^{-F_{\mu\tau}\cdot G_F^2 M_{Pl}T_{i,\mu\tau}^3}\,,\;\;\;
\Delta_{e}(T_d)=\Delta_{e}(T_{i,e})\e^{-F_e\cdot G_F^2 M_{Pl}T_{i,e}^3}\,.
\]
The oscillations involving electron neutrinos start later, and the limited time tends to make the asymmetry redistribution much less effective  compared to the case of two other neutrinos. The final results also depend on the neutrino sector parameters (on the values of the parameters $F_{\mu\tau}$, $F_e$ in our illustrative example). 

\begin{figure}[!htb]
    \centerline{
    \includegraphics[width=0.5\textwidth]{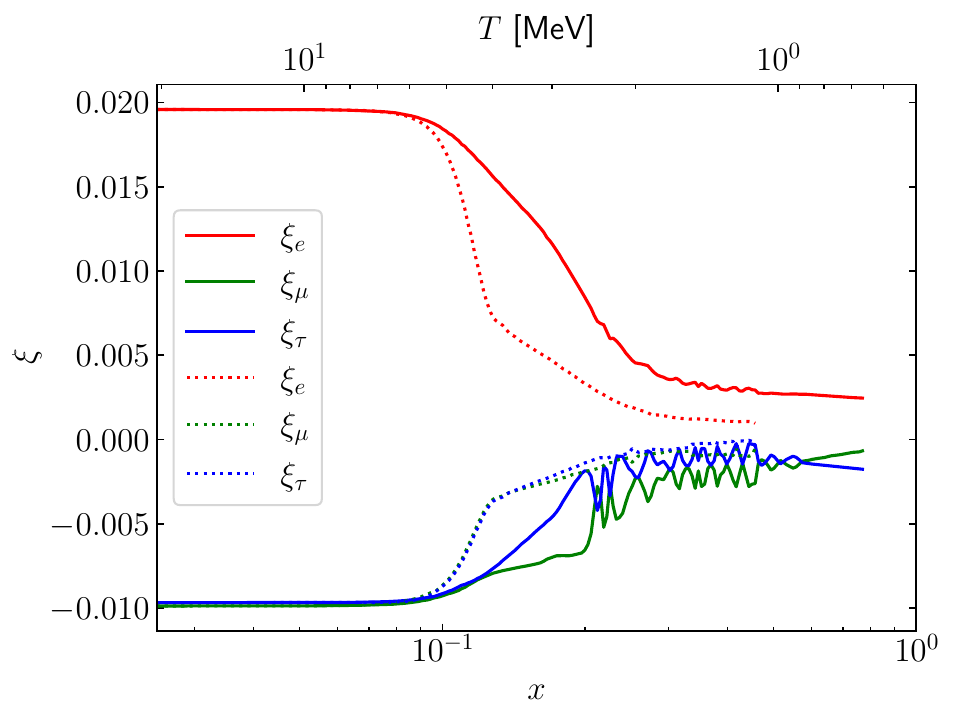}
\includegraphics[width=0.5\textwidth]{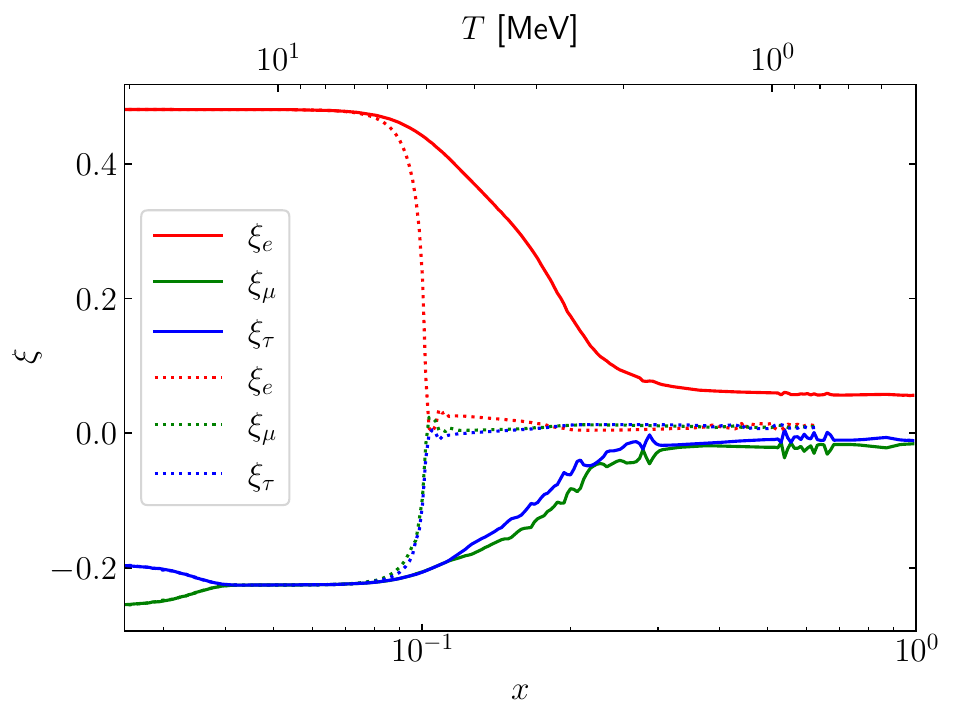}
    }
    \caption{Evolution of the neutrino asymmetries, electron neutrino (red lines), muon neutrino (green lines) and tau neutrino (blue lines) with plasma temperature, $x\equiv m_e/T$, at the oscillation  stage. The initial lepton asymmetries at 30\,MeV are $(5.97,-3.01,-2.95)\cdot 10^{-4}$ (left panel) and $(146.61, -78.29, -59.36)\cdot 10^{-4}$ (right panel). The neutrino oscillation parameters are taken from \cite{ParticleDataGroup:2024cfk} for normal (solid lines) and inverse (dotted lines) ordering of the neutrino masses.}
    \label{fig:asymmetry}
\end{figure}

In Fig.\,\ref{fig:asymmetry} 
we illustrate this chain of reasoning by numerical calculations of sterile neutrino production performed with 
\texttt{dmpheno} code \cite{Ghiglieri:2015jua} for 
$m_{\nu_s}=7.11$\,keV, mixing with electron neutrino, $\sin^22\theta=2.8\cdot 10^{-13}$, and two different sets of initial neutrino flavor asymmetries, $(\Delta_{\nu_e},\Delta_{\nu_\mu}, \Delta_{\nu_\tau})$, namely, $(2,-1,-1)\cdot 10^{-4}$ (left panel) and $(50,-25,-25)\cdot 10^{-4}$ (right panel) and subsequent evolution of the neutrino asymmetries at the stage of active neutrino oscillations, performed with \texttt{FortEPiaNO} code \cite{Gariazzo:2019gyi} and present estimates of the neutrino oscillation parameters\,\cite{ParticleDataGroup:2024cfk}. The values of interest are the ratios of the neutrino chemical potentials to the temperature $\xi_{\nu_\alpha}\equiv \mu_{\nu_\alpha}/T$. They are related to the neutrino asymmetries $\Delta_{\nu_\alpha}$ used above as follows, 
\[
\xi_{\nu_\alpha}=\Delta_{\nu_\alpha}\cdot \frac{4\pi^2}{15}h_{eff}\,,
\]
with $h_{eff}$ referring to the effective number of degrees of freedom in the cosmic plasma entering the entropy density of the interesting epoch. The sterile neutrino production for the chosen model parameters terminates at about $T=30$\,MeV, and from that moment we run the \texttt{FortEPiaNO} code with initial value of $h_{eff}(30\,\text{MeV})=12.47$ \cite{Ghiglieri:2015jua} and initial neutrino asymmetries $\Delta_{\nu_\alpha}$ (the outcome of the \texttt{dmpheno} running) equal 
to $(5.97,-3.01,-2.95)\cdot 10^{-4}$ (left panel) and $(146.61, -78.29, -59.36)\cdot 10^{-4}$ (right panel). We use linearly spaced 30-point momentum grid from $y_{min}=0.01$ to $y_{max} = 20$ with full momentum dependency. Such settings ensure a sufficient accuracy of the calculations. To perform this calculation, we modified the public version of the code \texttt{FortEPiaNO} by introducing the neutrino asymmetry matrix, which accounts for the neutrino asymmetry potential $V_\alpha^L$.

We find that the interesting electron neutrino asymmetry starts to reduce quite late indeed, but the efficiency of the asymmetry reduction differs in the two cases of mass hierarchy due to different combination of the mixing angles determining the numerical factor $F$ in the reduction exponent of \eqref{washing}.  In case of normal hierarchy (solid lines in Fig.\,\ref{fig:asymmetry}) {\it the reduction factor is only about ten.} Since numerically $\xi_{\nu_e}\approx 33 \Delta_{\nu_e}$, with the present limit from BBN $|\xi_{\nu_e}|\lesssim0.01$ \cite{ParticleDataGroup:2024cfk,Escudero:2022okz} one may use the initial electron neutrino asymmetry as large as $\Delta_{\nu_e}\simeq 10^{-3}$. In case of inverse hierarchy (dotted lines in Fig.\,\ref{fig:asymmetry}) the reduction factor is much larger, and one may adopt much higher initial neutrino asymmetry.  

\begin{figure}[!htb]
    \centerline{
    \includegraphics[width=0.5\textwidth]{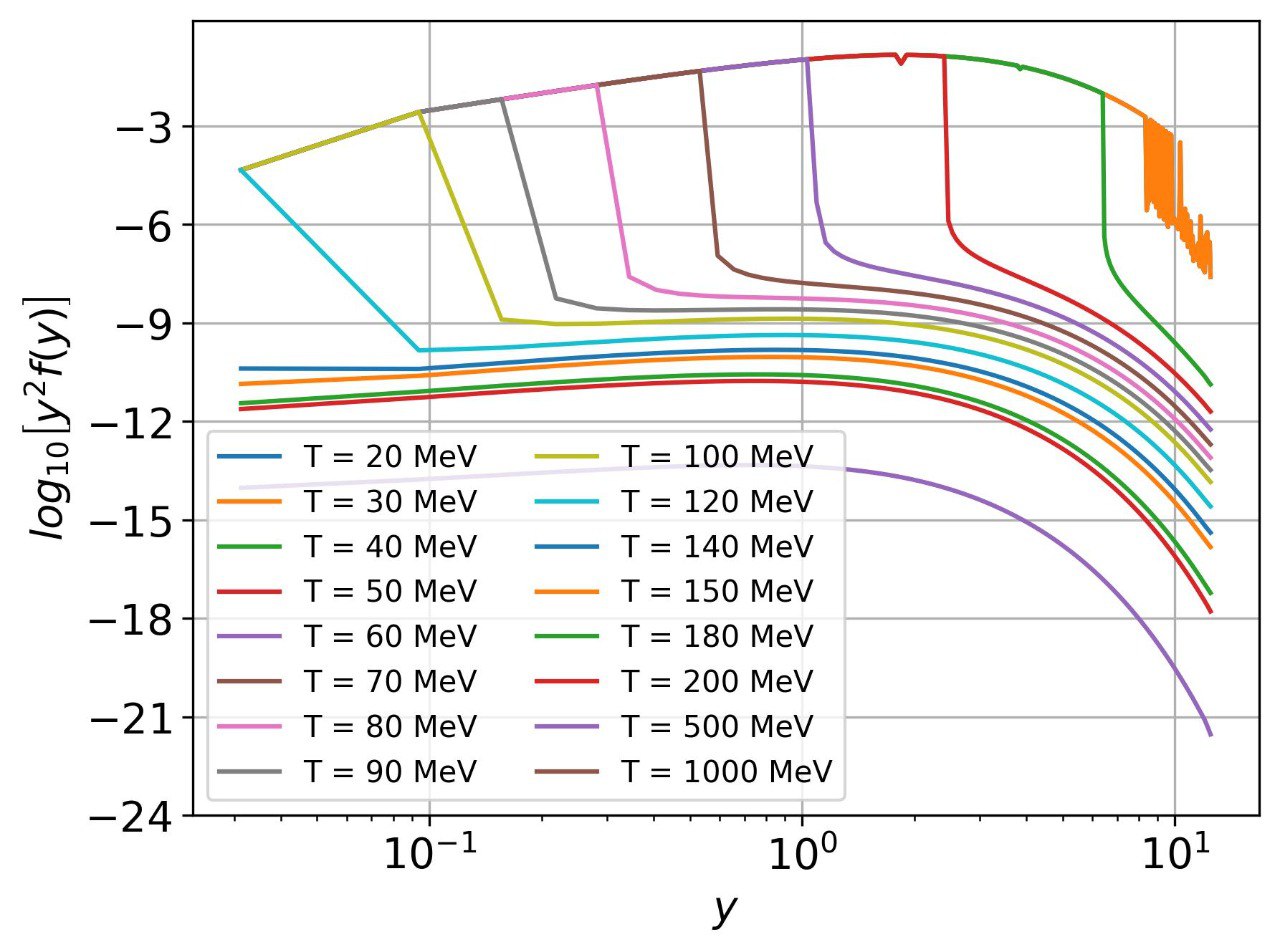}
\includegraphics[width=0.5\textwidth]{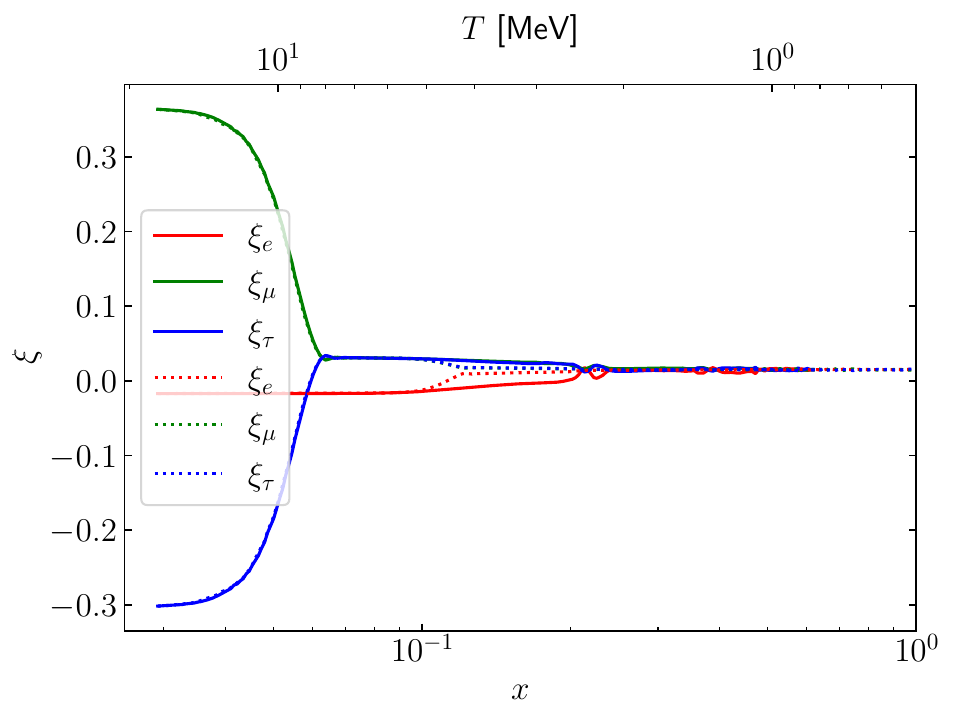}
    }
    \caption{Sterile neutrino spectrum at various temperatures (left panel) and subsequent evolution of the neutrino flavor asymmetries (right panel), electron neutrino (red lines), muon neutrino (green lines) and tau neutrino (blue lines) with plasma temperature, $x\equiv m_e/T$, at the oscillation  stage. The initial lepton asymmetries at 20\,MeV are $(-6.0208,128.8341,-106.8030)\cdot 10^{-4}$, sterile neutrino mass $m=7.11$\,keV, the {\it mixing with muon neutrino} is tuned to fully explain the dark matter component, $\sin^22\theta=7.36\cdot 10^{-14}$. The calculations are performed for the best fit parameters of the neutrino sector taken from Ref.\,\cite{ParticleDataGroup:2024cfk} for both normal (solid lines) and inverse (dotted lines) hierarchies.
    }
    \label{fig:alternative}
\end{figure}

{\bf 4.} So far we considered the sterile neutrino mixed with electron neutrino. There is more room for mixing with muon or tau neutrinos, where the initial electron neutrino asymmetry is close to zero, and muon and tau neutrino asymmetries are large, equal in value but of opposite sign. Such initial conditions may be obtained with specific symmetry in the sector responsible for generation of the initial lepton asymmetry in the primordial plasma (possibly after electroweak transition). As we discussed above, the active neutrino oscillations between muon and tau neutrinos start earlier, at 3 times higher temperature, and the cancellation of the total asymmetry in this case is much more efficient. In Fig.\,\ref{fig:alternative} 
we illustrate this point by numerical calculations performed for the mixing with muon neutrino, $\sin^22\theta=1.2\cdot 10^{-13}$, $m_{\nu_s}=7.11$\,keV and the initial set of neutrino asymmetries $(0,45,-45)\cdot 10^{-4}$, which evolves to $(-6.0208,128.8341,-106.8030)\cdot 10^{-4}$ by the onset of active neutrino oscillations. As we discussed above, see Fig.\,\ref{fig:spectra}, with higher lepton asymmetry the sterile neutrino production terminates later, so we run  the \texttt{FortEPiaNO} code from $T=20$\,MeV when $h_{eff}=11.28$. 
One observes that the muon and tau lepton asymmetries equalize very rapidly and neatly, as expected, reaching values $309.90\cdot 10^{-4}$ and $309.93\cdot 10^{-4}$ at $T=6.9$\,MeV. We find also that the  asymmetry dynamics is somewhat different in the cases of normal and inverse mass hierarchy, see right panel of Fig.\,\ref{fig:alternative}.  

Note, that while we start with zero total neutrino asymmetry, it becomes non-zero after the sterile neutrino generation, since the asymmetry, mostly from the flavor which mixes with sterile neutrinos, gets transferred to the sterile neutrinos. Some asymmetry also emerges in active neutrinos, which do not mix with the sterile component. In principle, one can tune the initial pattern of neutrino flavor asymmetry in such a way, that after termination of the sterile neutrino production the electron neutrino asymmetry equals zero. It requires some small but non-zero initial asymmetry in the electron neutrinos. 
Likewise, one can start with a set of sterile neutrino asymmetries, when after production both the electron neutrino and the total lepton asymmetries are zero. 
Whether such tunings may be adopted within concrete mechanisms capable of generating the neutrino asymmetries in the early Universe remains to be explored. It is worth mentioning that with a mass in the keV range the sterile neutrino itself contributes to the energy density at BBN. Its number density at present is of about 1/cm$^{3}$ or lower. Comparing it to about 100/cm$^{3}$, the present number density of the active neutrino specie, see e.g.\,\cite{Rubakov:2017xzr}, one concludes that the sterile neutrino contribution to the energy density in the BBN epoch is less than one per cent of that of a single active neutrino specie, and hence is negligible, given the present and near future accuracy expected in measurements of the primordial abundances of the light chemical elements, see e.g.\,\cite{ParticleDataGroup:2024cfk}.

{\bf 5.} To summarize, in this letter we show that the presently allowed model parameter space of the sterile neutrino dark matter may be significantly extended with specific tuned 
neutrino flavor asymmetries. If large in magnitude, but the sum is close to zero, it can be efficiently canceled in the subsequent oscillations of the active neutrino. The inverse ordering of neutrino masses shows more efficient cancellation. Hence, at the moment of neutron freeze-out the electron asymmetry may be greatly reduced, even to zero. It mitigates the limits from primordial nucleosynthesis, allows for higher initial lepton asymmetry and hence lower sterile-neutrino mixing angles. Likewise, with a higher initial lepton asymmetry the resonant mechanism produces colder sterile neutrinos. We show, that if the lepton asymmetry disappears from the plasma, the resulting spectrum is indeed colder. At the same time, higher asymmetry postpones the sterile neutrino production, and with sufficiently high asymmetry the active neutrino oscillations may do the job, cancel the asymmetry and make the sterile neutrino spectrum cold.     

Note in passing that some electron asymmetry at the level of few percent may be welcome given some anomalous abundances of primordial elements recently observed, see e.g.\cite{Burns:2022hkq,Escudero:2022okz}.   

It is worth mentioning that while there are other numerical codes to calculate the neutrino oscillations in the primordial plasma, we believe that the qualitative results do not depend on our choice. The qualitative picture with various ways of reopening the areas of the model parameter space is correct. 

\begin{figure}[!htb]
    \centerline{
    \includegraphics[width=0.5\textwidth]{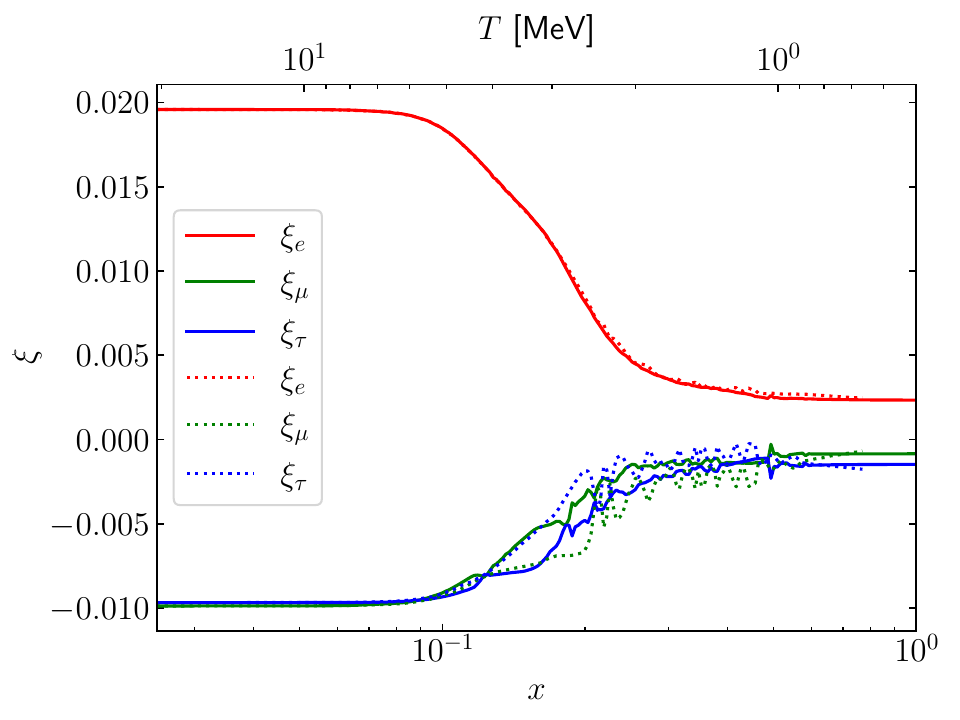}
\includegraphics[width=0.5\textwidth]{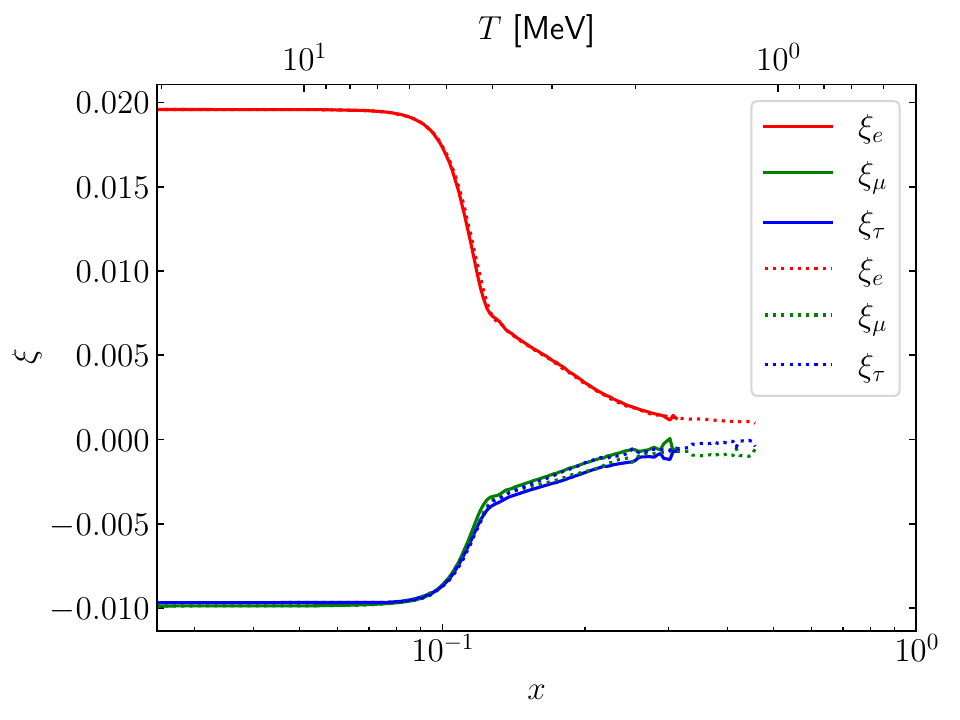}
    }
    \caption{Evolution of the neutrino flavor asymmetries with plasma temperature, $x\equiv m_e/T$, for the normal hierarchy (left panel), and inverse hierarchy (right panel), with taken from \cite{ParticleDataGroup:2024cfk} values of CP-phase (dotted lines) and zero CP-phase (solid lines). The initial pattern of the lepton asymmetries for both plots is the same as for the left plot in Fig.\,\ref{fig:asymmetry}. 
    }
    \label{fig:misc}
\end{figure}

Without using a particular mechanism to produce the original lepton asymmetry, one is quite free to tune the parameters in the sterile neutrino sector. We illustrate our statements with numerical calculations performed for particular values of sterile neutrino mass and mixing angle. The parameters of the active neutrino sector are fixed by neutrino oscillation experiments, except the mass hierarchy and the value of CP-phase. As we observed above, the inverted hierarchy is capable of reducing the net neutrino asymmetry more efficiently. We fix the value of CP-phase as suggested in \cite{ParticleDataGroup:2024cfk}, but the present accuracy of it's determination is poor. Thus we run  the \texttt{FortEPiaNO} code with the suggested value of CP-phase and with zero CP-phase for the set of neutrino asymmetries and do not find any significant change in the dynamics of asymmetry, see 
Fig.\,\ref{fig:misc},  
that is consistent with detailed studies\,\cite{Gava:2010kz,Froustey:2021azz} of the impact of the CP-phase on the evolution of neutrino flavor asymmetries.

\vskip 0.3cm
{\it Note added.} When the paper was finished we observed new work \cite{Domcke:2025lzg}, where the lepton asymmetry evolution due to active neutrino oscillations in the primordial plasma was considered. Their results on the efficiency of inverse hierarchy and inefficiency of CP-phase are consistent with ours.     

\vskip 0.3cm

We thank M.\,Shaposhnikov for very valuable correspondence and discussions which at some time have been transformed into this  project. 
The authors thank S.\,Gariazzo for valuable correspondence related to the  \texttt{FortEPiaNO} operation.   
This work is supported in the framework of the State project ``Science'' by the Ministry of Science and Higher Education of the Russian Federation under the contract 075-15-2024-541. DK thanks the Foundation for the Advancement of Theoretical Physics and Mathematics “BASIS” for the individual grant.


\vskip 1cm

\bibliographystyle{utphys}
\bibliography{refs}
\end{document}